\documentclass[journal,12pt,onecolumn,draftclsnofoot,]{IEEEtran}
\usepackage{amsmath,amsfonts}
\usepackage{amssymb}
\usepackage{array}
\usepackage[caption=false,font=normalsize,labelfont=sf,textfont=sf]{subfig}
\usepackage{textcomp}
\usepackage{stfloats}
\usepackage{url}
\usepackage{verbatim}
\usepackage{graphicx}
\usepackage{comment}
\usepackage{cite}
\usepackage{xcolor}
\usepackage{algorithm}
\usepackage{subcaption}
\usepackage{hyperref}
\usepackage{caption}
\usepackage{algorithmicx}
\usepackage{algpseudocode}
\def\BibTeX{{\rm B\kern-.05em{\sc i\kern-.025em b}\kern-.08em
    T\kern-.1667em\lower.7ex\hbox{E}\kern-.125emX}}
\usepackage{balance}
\begin{document}
\title{Investigation of RIS Integration in Satellite Systems Guided by ITU-R VSAT Radiation Diagram}

\author{Symeon~Vaidanis,~\IEEEmembership{Graduate Student Member,~IEEE,}
        and~Athanasios~D.~Panagopoulos,~\IEEEmembership{Senior Member,~IEEE}
\thanks{S. Vaidanis is with the Communication Systems Department of EURECOM, Sophia Antipolis, 06410 France, and A. D. Panagopoulos is with the School of Electrical and Computer Engineering of National Technical University of Athens, 15780 Greece. E-mail: (see symeon.vaidanis@eurecom.fr, thpanag@ece.ntua.gr). Symeon Vaidanis is supported by the Onassis Foundation - Scholarship ID: F ZU 076-1/2024-2025.}}

\maketitle

\begin{abstract}
The introduction of Reconfigurable Intelligent Surfaces (RIS) technology marks a transformative era in modern communication systems, enhancing coverage and facilitating higher data transmission rates. While recent research has delved into the integration of RIS into satellite networks, a critical and practical gap persists concerning its implementation in the satellite-Very Small Aperture Terminal (VSAT) downlink, specifically amplifying contributions from the VSAT side. This paper addresses this gap by proposing a simple and engineering methodology to optimize RIS position and orientation, incorporating considerations for the antennas' radiation patterns and the system's topology to streamline optimization variables. The primary objective of this paper is to alleviate the double fading constraint in the reflect link without the need for active elements, achieved by leveraging the Fresnel zone of the RIS. The improvement power ratio depends on the frequency and the VSAT 's diameter. The results show that for a VSAT with diameter 1m the gain is up to 9dB at 2 GHz (S band) and up to 14dB at 4 GHz (C band). Additionally, the RIS channel equation is refined, encompassing the tropospheric attenuation and accounting for electromagnetic wave polarization.
\end{abstract}

\begin{IEEEkeywords}
Position and orientation optimization, reconfigurable intelligent surfaces, satellite networks
\end{IEEEkeywords}

\section{Introduction}
Reconfigurable Intelligent Surfaces (RIS) have emerged as a transformative technology poised to enhance modern communication systems, promising expanded coverage and elevated data transmission rates \cite{RIS_Survey_1, RIS_Survey_2}. In recent years, the scientific community has turned its attention to exploring the integration of RIS in satellite networks \cite{RIS_Satellite_Survey_1, RIS_Satellite_Survey_2}. The current bibliography unveils a spectrum of applications, each contributing to the advancement of communication systems. Noteworthy among these applications is the proposal in \cite{RIS_Satellite_Hopping}, where RIS is strategically employed on neighboring Low Earth Orbit (LEO) satellites operating in the Terahertz (THz) band. This deployment aims to alleviate misalignment fading, compensate for high path loss, and consequently enhance the Signal-to-Noise Ratio (SNR). Investigating scenarios devoid of Line of Sight (LoS) paths between User Equipment (UE) and satellites, papers \cite{RIS_Satellite_NLoS} and \cite{RIS_Satellite_Refracting} offer distinct perspectives. The former focuses on minimizing total transmit power while meeting UE rate requirements, while the latter proposes innovative deployments to extend the coverage of RIS-aided links. Additionally, the discrete phase shift RIS takes center stage in \cite{RIS_Orientation_Optimization}, studying its orientation in a cellular environment without analytically defining the orientation matrix. In paper \cite{RIS_Orinetation_Satellite}, authors solve the orientation and beamforming problem for the implementation of RIS in satellite networks. However, the topology of the RIS and the radiation patterns of ground terminals are ignored. The implementation of RIS in near-field conditions within IoT non-terrestrial networks is explored in \cite{RIS_Sat_Near_Field}, and in \cite{RIS_LEO_Mobility}, RIS is employed to mitigate the Doppler effect in Low Earth Orbit (LEO) satellites. In addition, RISs are proposed to provide a solution to the interference mitigation between LEO and Geostationary (GEO) satellites in \cite{RIS_LEO_GEO}. In paper \cite{RIS_Multisatellite}, authors optimize the phase shift of the RIS and the beamforming vector of UE in order to perform in a multi-satellite cooperative system. Finally, authors in \cite{RIS_Satellite_NOMA} explore the implementation of RIS to satellite–terrestrial integrated networks with Non-Orthogonal Multiple Access (NOMA); and in \cite{RIS_Multicast_Satellite}, they propose simultaneously transmitting and reflecting RIS (STAR-RIS) for satellite communication.

Building upon the foundations laid by the current bibliography, this paper introduces several key contributions. Firstly, this work focuses on optimizing the position and orientation of RIS in relation to a ground terminal VSAT in a satellite network. The radiation patterns of the VSAT and a RIS element play a significant role to an achievable improvement and also determine the topology which achieves this maximum achievable improvement. In addition, a comparative analysis is conducted, investigating the implications of directing the VSAT beam towards the satellite versus the RIS. The current design direction of these terrestrial satellite networks is to orient the VSAT towards the satellite and use cooperative reception or adaptive transmission techniques to improve the power margin of the link. In this study, we propose the VSAT to be oriented towards the RIS, which can be considered as an additional parabolic antenna with the property of both mechanical and electronical rotation of radiation pattern. Furthermore, the challenge of double fading in the reflect path is addressed by leveraging the Fresnel zone of the RIS, without requiring active elements. The presence of active elements is following by the demand of additional power consumption and more complex electronic hardware, and thus the main advantages of a passive RIS, which are the low energy consumption and the low-cost manufacturing, retreat. Finally, the RIS channel equation is enhanced by incorporating tropospheric attenuation and accounting for the polarization of electromagnetic waves, guided by ITU-R. These propagation phenomena are extremely important for satellite communications and affect their performance.

\section{System Model}
\subsection{Antennas theory preliminaries and radiation zones}
To start with the antennas' radiation patterns, the antenna mask for VSAT complies with guidelines from the International Telecommunication Union's Radiocommunication Sector (ITU-R) \cite{ITU_R_FSS},\cite{FSS_Radiation_Pattern}, specifically designed for small ground terminals operating as Fixed Satellite Systems (FSS) across the frequency bands from 2 GHz to 31 GHz. The VSAT's directional gain is symmetric to the direction of the beam focus, exhibiting cylindrical symmetry.The far field formula for a parabolic antenna is 
\begin{equation}
    R_\text{far field}^\text{parabolic} = \frac{2 \cdot D^2}{\lambda}
\end{equation}
where $D$ is the diameter of the parabolic antenna and $\lambda$ is the wavelength.  

The directional gain of the RIS element, outlined in paper \cite{RIS_Channel_Modeling_1}, is symmetric to the direction of beam focus (cylindrical symmetry), independent of the operating frequency, and depends only on the patch size, which is defined only for the half space. The far field formula for a patch antenna  is 
\begin{equation}
    R_\text{far field}^\text{patch} = \frac{2 \cdot (d_x^2 + d_y^2)}{\lambda}
\end{equation} 
where $d_x$ and $d_y$ are the two dimensions of the patch antenna. The far field formula for a rectangular antenna array is 
\begin{equation}
R_\text{far field}^\text{array} = \frac{2 \cdot (a^2 + b^2)}{\lambda}
\end{equation} 
where $a$ and $b$ is the two dimensions of the rectangular antenna and the radiating near field of it is 
\begin{equation}
R_\text{radiating near field}^\text{array} = 0.62 \cdot \sqrt{\frac{(a^2 + b^2)^{\frac{3}{2}}}{\lambda}}
\end{equation}

\begin{figure}
    \centering
    \includegraphics[width=0.6\linewidth]{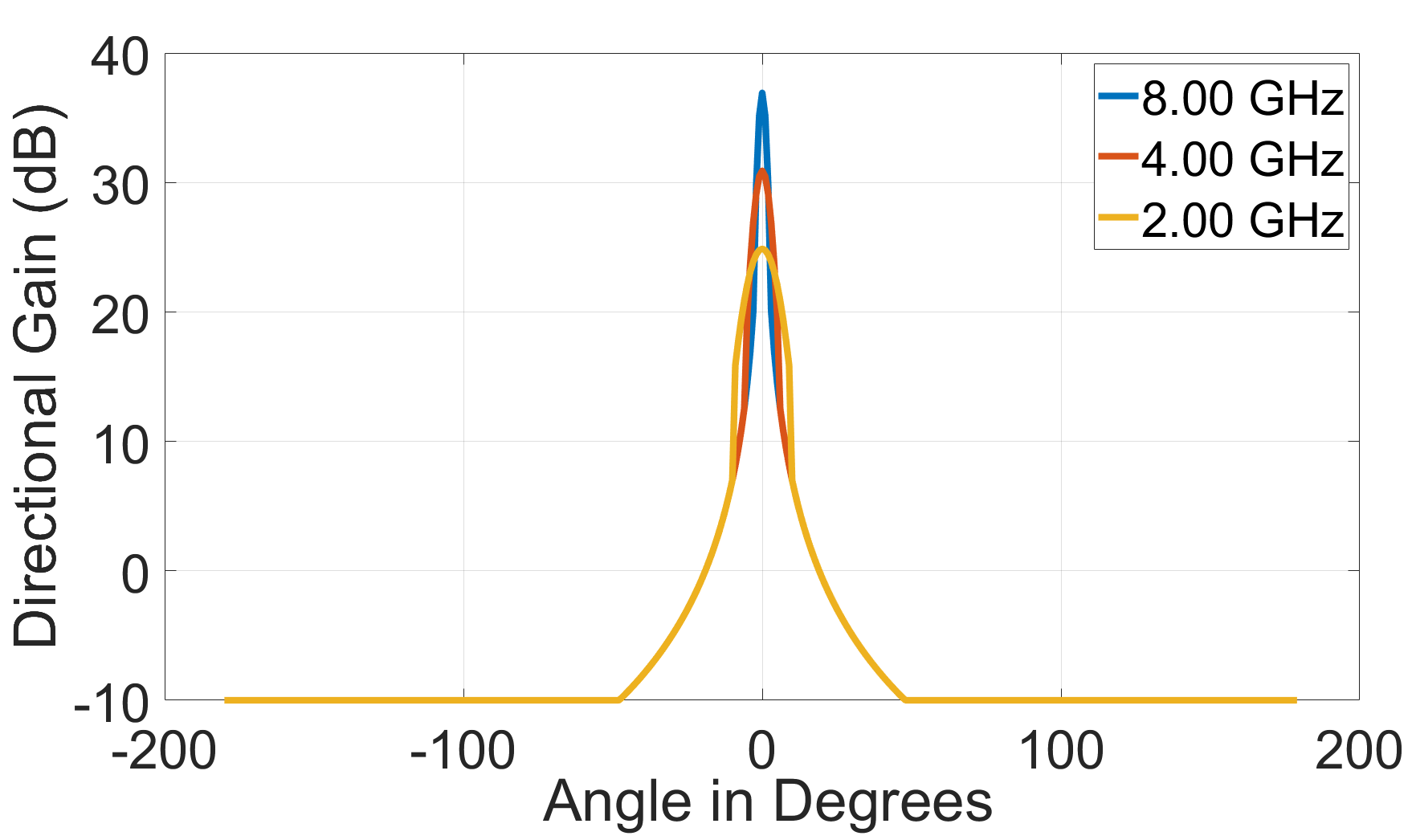}
    \caption{Radiation diagrams of VSAT (ITU-R 456-6), D = 1m}
    \label{fig:rad.diagrams.VSAT.1}
\end{figure}
\begin{figure}
    \centering
    \includegraphics[width=0.6\linewidth]{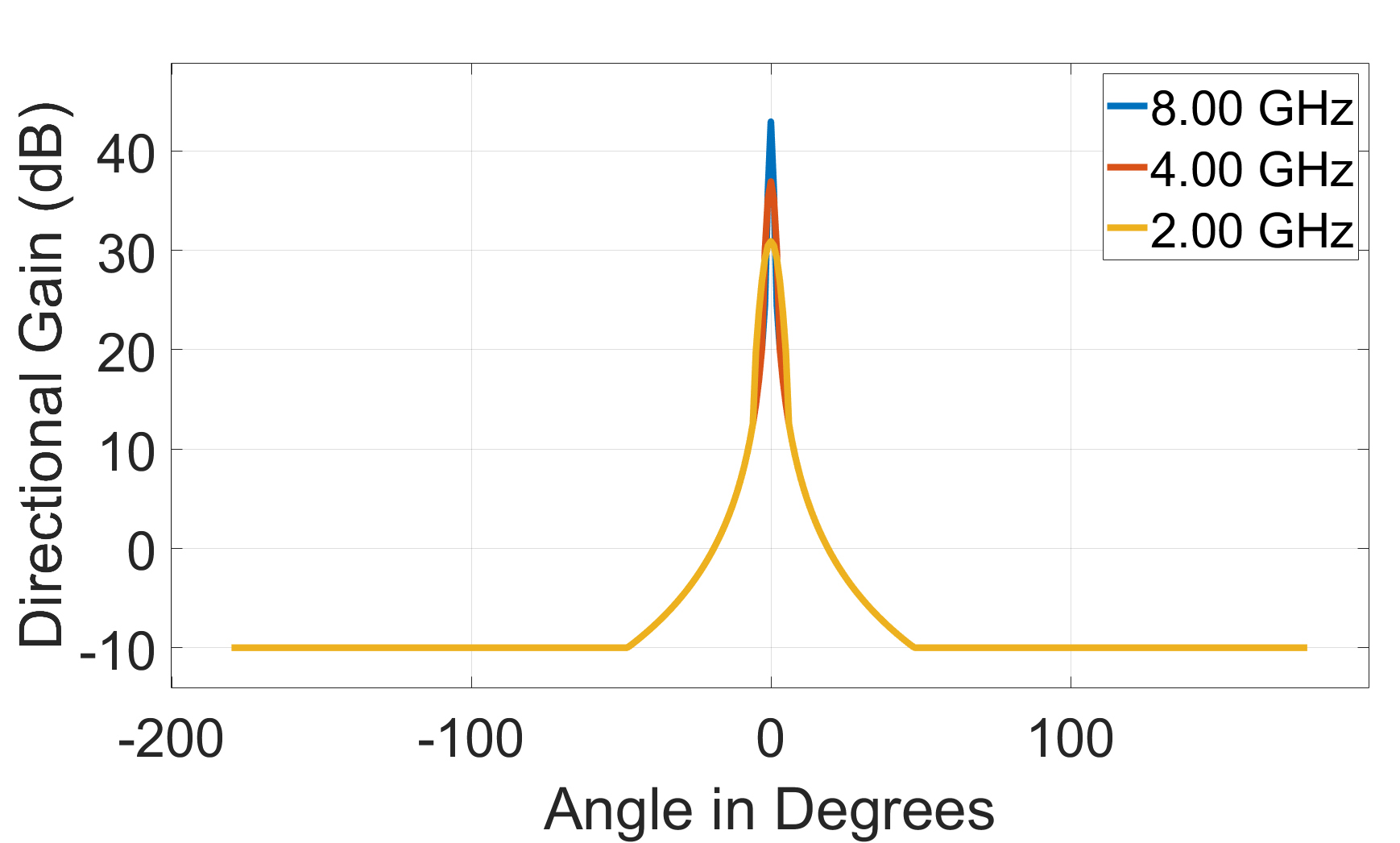}
    \caption{Radiation diagrams of VSAT (ITU-R 456-6), D = 2m}
    \label{fig:rad.diagrams.VSAT.2}
\end{figure}
\begin{figure}
    \centering
    \includegraphics[width=0.6\linewidth]{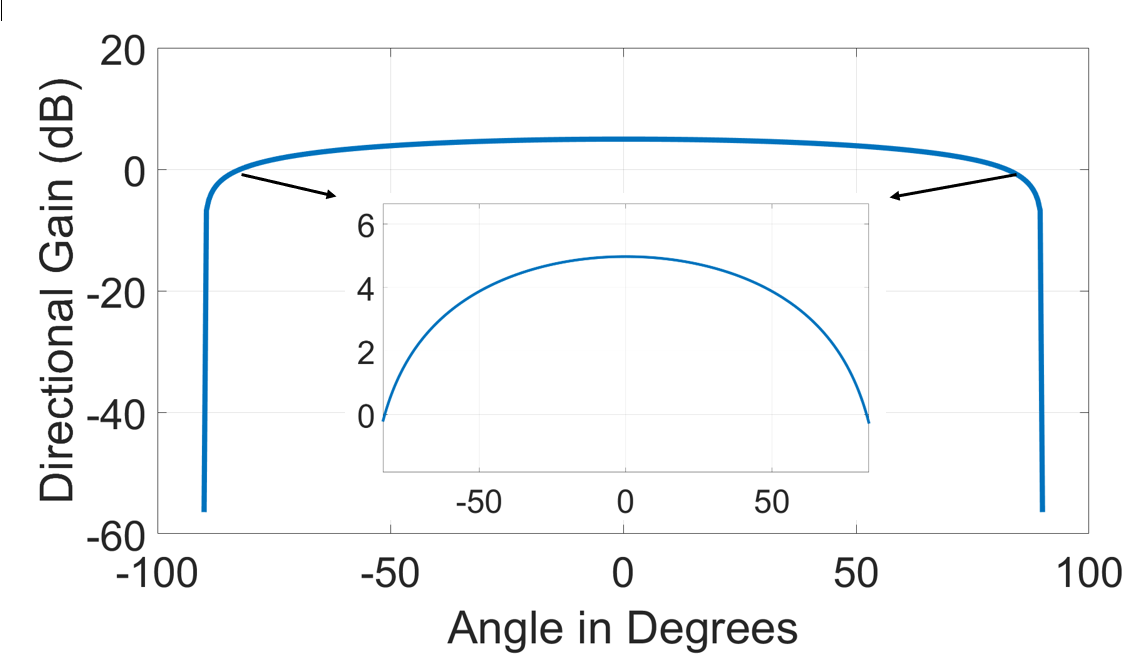}
    \caption{Radiation Diagrams of RIS element}
    \label{fig:rad.diagrams.RIS}
\end{figure}

Defining the radiation zones requires addressing critical distances and their relationships.  The distance between VSAT and the satellite is calculated using the formula \cite{Ha}. In this book, the formula of the elevation angle of the satellite is determined, involving the parameters of Earth's radius, satellite height, VSAT's longitude and latitude, and the satellite's latitude. Also, cooperation between satellite networks and terrestrial systems is analyzed in \cite{Sak}. 

In addition, the physical dimensions of the examined system, independent of the type of satellite system technical characteristics, give us the ability to consider 
\begin{equation}
    d_{sat}^{RIS} \gg d_{RIS}^{vsat} \text{ and } d_{sat}^{RIS} \approx d_{sat}^{vsat}
\end{equation} 
Further, because the directional gain mask of VSAT and the directional gain of the RIS element have been defined for the far field for each one, the distance between RIS and the VSAT should be $d_{RIS}^{vsat} \geq R_\text{far field}^\text{parabolic}$, as in the majority of the cases $R_\text{far field}^\text{parabolic} \gg R_\text{far field}^\text{patch}$. Moreover, the selection of the appropriate channel model based on the topology and the engineering restrictions of the system is an issue of paramount importance. 
\begin{itemize}
    \item If the inequality $R_\text{far field}^\text{RIS} < R_\text{far field}^\text{parabolic}$ is satisfied, the far field approximation can be used.
    \item On the contrary, if the inequality $ R_\text{radiating near field}^\text{RIS} < R_\text{far field}^\text{parabolic} < R_\text{far field}^\text{RIS}$ is satisfied and at the same time $d_{RIS}^{vsat} \leq R_\text{far field}^\text{RIS}$, the analytical near field equations should be used.
\end{itemize}
For the case of a rectangular RIS ($M=N$) and rectangular elements ($d_x = d_y$) with value $d_x = \frac{\lambda}{w},w\in\mathbb{N}^*$, the expression between the far field bounds of VSAT and the RIS can be transformed into expression between the sizes of VSAT and RIS as follows:
\begin{subequations}
    \begin{equation}
        \label{eq: Far Field Approximation Bound}
        M < \frac{w}{\sqrt{2}}\frac{D}{\lambda} 
    \end{equation}
    \begin{equation}
        \label{eq: Freznel Zone Bound}
        \frac{w}{\sqrt{2}}\frac{D}{\lambda}  < M < 1.54w \left(\frac{D}{\lambda}\right)^{\frac{4}{3}}
    \end{equation}
    \label{eq:Near and Far Field bounds}
\end{subequations}
where (\ref{eq: Far Field Approximation Bound}) one describes the case, in which we can use far field approximation, while (\ref{eq: Freznel Zone Bound}) illustrates the case, in which we might have to use an analytical near field model.

\begin{table}
    \centering
    \caption{ The parameters of the optimization problem}
    \begin{tabular}{|m{1.1cm}|m{5cm}|}
        \hline
        $d_{sat}^{vsat}$ & Distance between satellite and VSAT\\
        \hline
        $d_{sat}^{RIS}$ & Distance between satellite and RIS\\
        \hline
        $d_{RIS}^{vsat}$ & Distance between RIS and VSAT\\
        \hline
        $G_{sat}^{RIS,(m,n)}$ & Directional gain of satellite towards the $(m,n)$ element of RIS\\
        \hline
        $G_{RIS,(m,n)}^{sat}$ & Directional gain of the $(m,n)$ element of RIS towards the satellite\\
        \hline
        $G_{RIS,(m,n)}^{vsat}$ &  Directional gain of the $(m,n)$ element of RIS towards the VSAT\\
        \hline
        $G_{vsat}^{RIS,(m,n)}$ & Directional gain of VSAT towards the $(m,n)$ element of RIS\\
        \hline
        $\hat{\rho}_{sat}$ & The unity vector of polarization of satellite\\
        \hline
        $\hat{\rho}_{VSAT}$ & The unity vector of polarization of VSAT\\
        \hline
        $\hat{\rho}_{RIS}$ & The unity vector of polarization of RIS\\
        \hline
        $A_{m,n}$ & The amplitude of reflection coefficient of the $(m,n)$ element of RIS\\
        \hline
    \end{tabular}
    \label{tab:my_label}
\end{table}

\subsection{Coordination systems}
The main direction of optimizing the position of RIS leads us to define the topology of the system using a Global Coordination System (GCS) as reference and two Local Coordination Systems (LCS), one for VSAT (LCS-VSAT) and one for RIS (LCS-RIS). Each of the LCSs can be described by the position vector ($\text{center}_{\text{LCS}}$) of their center with reference to the GCS and the rotation vector ($\text{rotation}_{\text{LCS}}$), which includes the bearing angle, the down tilt angle and the slant angle and describes the orientation of the radiation pattern of the antenna. More precisely, the bearing ($\alpha$), down tilt ($\beta$) and slant ($\gamma$) angles refer to the anti-clockwise rotation following the right hand rule around the $z$, $y$ and $x$ axis respectively \cite{3GPP_5G_channel_modeling}. 
\begin{figure}
    \centering
    \includegraphics[width=0.6\linewidth]{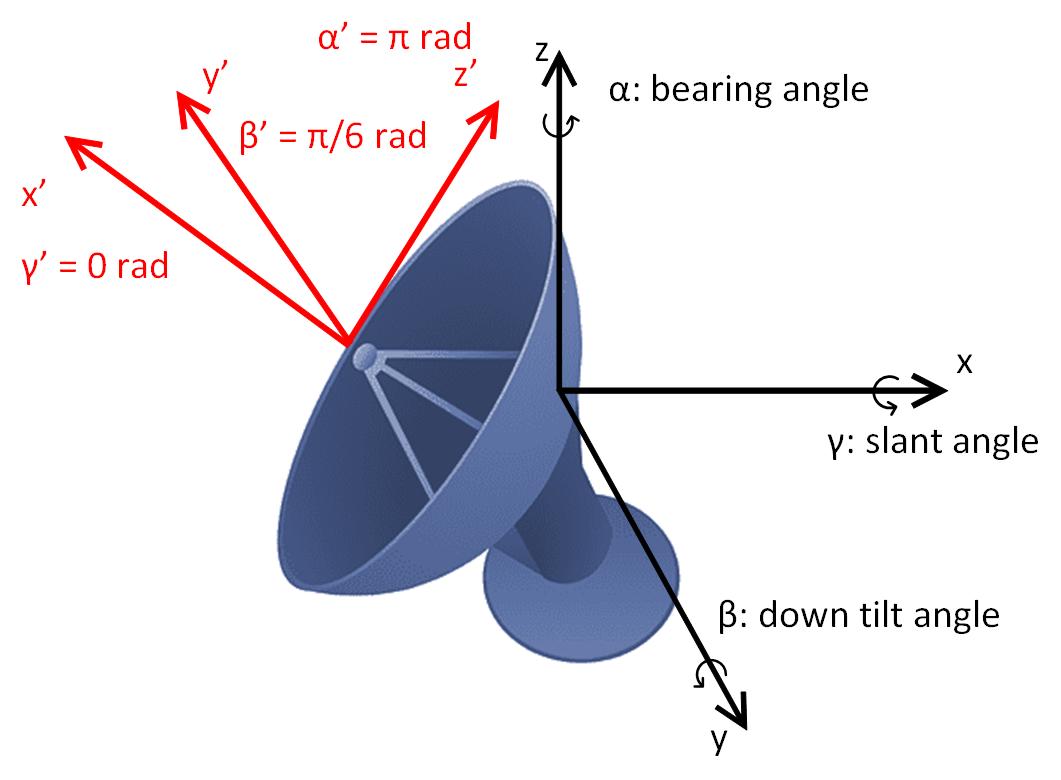}
    \caption{An example of rotation of GCS (black) to the LCS of VSAT (red). The $y$ axis is vertical to the surface of the paper.}
    \label{fig:VSAT_angle_parameters}
\end{figure}

\subsection{Path loss modeling}
The modeling of channel equation by using RIS has been proposed in \cite{RIS_Channel_Modeling_1} and can be considered as the superposition of links between a transmitter and receiver by using one element of the RIS independently from the others. The authors formulate a channel equation for distances larger than the bound of Fresnel distance of RIS and a simpler approximation of this equation, which can be applied to the far field of RIS. In this work, we focus only on a beamforming case, in which RIS is associated with only one VSAT. The multiple user perspective of this problem will be the main scope of future work. So, the channel coefficients for the beamforming case are the following:
\begin{subequations}
    \begin{equation}
    \label{eq: reflect}
        h_{refl}^{beam} = |\hat{\rho}_{sat} \cdot \hat{\rho}_{RIS}| \cdot |\hat{\rho}_{RIS} \cdot \hat{\rho}_{VSAT}| \cdot \sum _{m=1}^{M} \sum _{n=1}^{N} \sqrt{\frac{F_{m,n}}{L_{refl,(m,n)}}} A_{m,n}  
    \end{equation}
    \begin{equation}
    \label{eq: direct}
        h_{dir}^{beam} = | \hat{\rho}_{sat} \cdot \hat{\rho}_{VSAT} | \cdot \sqrt{\frac{G_{sat}^{vsat} G_{vsat}^{sat}}{L_{dir}}} 
    \end{equation}
\end{subequations} 
where (\ref{eq: reflect}) refers to the reflect path and (\ref{eq: direct}) to the direct path between the VSAT and a satellite. The component
\begin{equation}
    F_{m,n} = G_{sat}^{RIS,(m,n)} G_{RIS,(m,n)}^{sat} G_{RIS,(m,n)}^{vsat} G_{vsat}^{RIS,(m,n)}
\end{equation}
includes the directional gain of the satellite towards the $(m,n)$ element of RIS, the directional gain of $(m,n)$ element of RIS towards the satellite, the directional gain of $(m,n)$ element of RIS towards the VSAT and the directional gain of the VSAT towards the $(m,n)$ element of RIS. $G_{sat}^{vsat}$ is the directional gain of the satellite towards the VSAT and $G_{vsat}^{sat}$ is the directional gain of the VSAT towards the satellite. It should be mentioned that due to the fact that 
\[
    d_{sat}^{RIS} \approx d_{sat}^{vsat} \; \text{and} \; d_{sat}^{RIS} \gg d_{RIS}^{vsat} \Rightarrow 
\]
\begin{equation}
    G_{sat}^{RIS,(m,n)} \approx G_{sat}^{vsat}, \forall (m,n)
\end{equation}
These assumptions are established on the scale of the distances between the telecommunication objectives as $d_{RIS}^{vsat}$ has a value at the scale of 10m - 100m and $d_{sat}^{RIS}, d_{sat}^{vsat}$ at the scale of 10.000 km, depending on the type of a satellite system. Furthermore, 
\begin{equation}
L_{refl,(m,n)} = L^{trop}_{refl} \left(\frac{4\pi d_{sat}^{(m,n)}}{\lambda}\right)^2 \left(\frac{4\pi d_{(m,n)}^{vsat}}{\lambda}\right)^2 
\end{equation} 
and 
\begin{equation} 
L_{dir} = L^{trop}_{dir} \left(\frac{4\pi d_{sat}^{vsat}}{\lambda}\right)^2 
\end{equation} 
are the components which include the free space path loss and the attenuations due to troposphere $L^{trop}$   \cite{rain_attenuation,gases_attenuation,earth_space_propagation}, for the reflect and direct path respectively. 
As we study a beamforming case for one user, the large scale fading can be incorporated into the upper mentioned expression as another component and the small scale fading due to the propagation environment are supposed to be handle by perfect channel state information (CSI). 
Each one of the inner products' components in the expressions of channel coefficients describes the polarization alignment between the RIS and the satellite $|\hat{\rho}_{sat} \cdot \hat{\rho}_{RIS}|$, the RIS and the VSAT $|\hat{\rho}_{RIS} \cdot \hat{\rho}_{VSAT}|$, and the VSAT and the satellite $|\hat{\rho}_{sat} \cdot \hat{\rho}_{VSAT}|$. These parameters are very important for satellite communications due to the phenomenon of troposheric depolarization at higher frequencies. Finally, $A_{m,n}$ is the amplitude of the reflection coefficient of $(m,n)$ element of RIS. The impact of the space-time variations of propagation phenomena in the the path loss evaluation and in the interference is a subject of future work.

\section{Topology and Optimization of RIS Position and Orientation}
\begin{figure}
    \centering
    \includegraphics[width=0.5\linewidth]{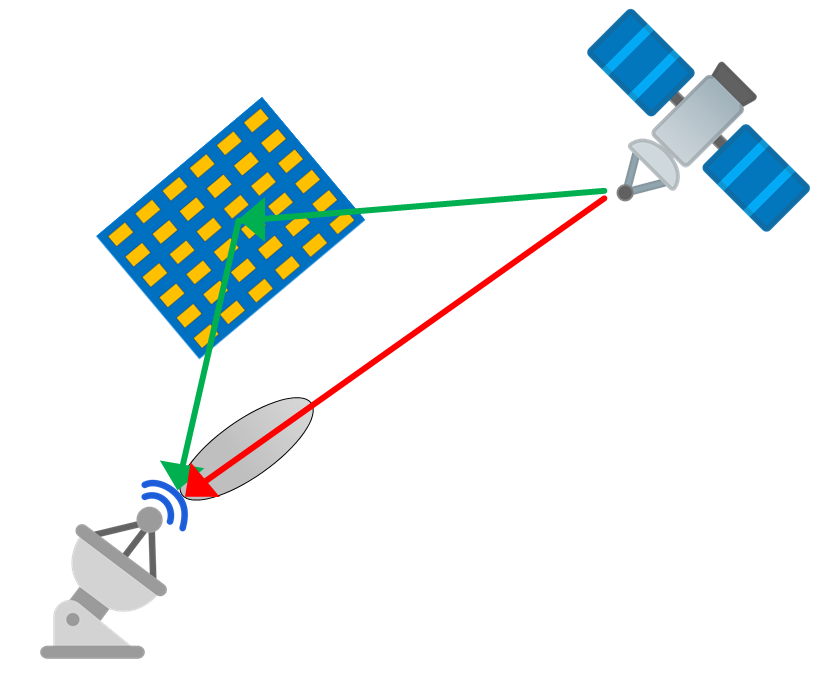}
    \caption{System topology for RIS-assisted in 2-D}
    \label{fig:RIS-assisted System Topology Primary in 2-D}
\end{figure}

\begin{figure}
    \centering
    \includegraphics[width=0.5\linewidth]{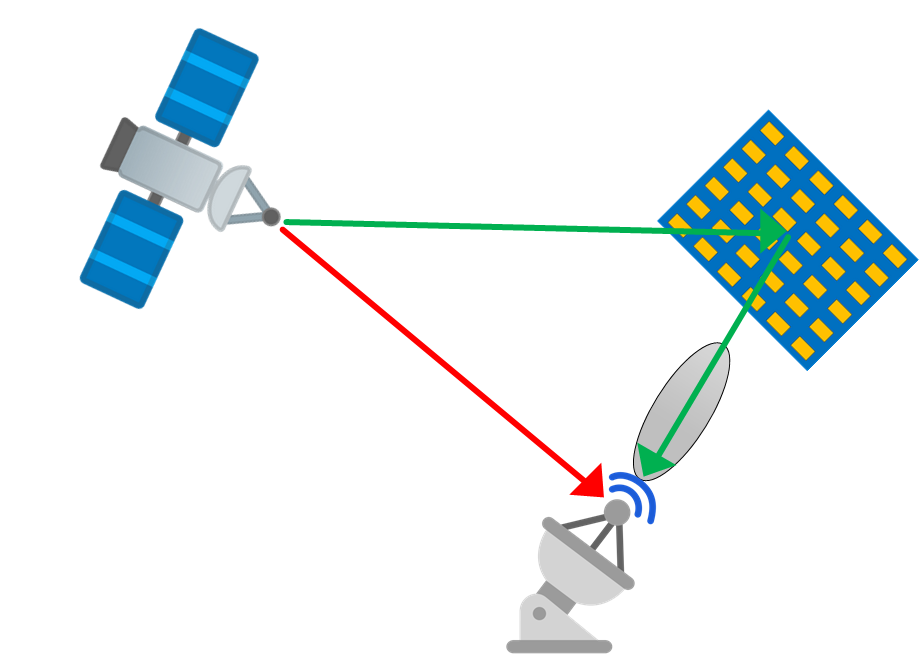}
    \caption{System topology for RIS-primary in 2-D}
    \label{fig:RIS-primary System Topology Primary in 2-D}
\end{figure}

When considering the LCS-VSAT, there are two main cases. Firstly, the VSAT can be rotated towards the satellite in order to align with it and as a result the RIS is going to add a reflection path to the standard direct path of transmission. It should be noticed that the location and the rotation of VSAT in this case is predefined and unchanged despite the presence of RIS. In the second case, the VSAT can be rotated towards the center of RIS and in this case the transmission path consists only of the reflection path. In the rest of the work, the first case will be refereed to as RIS-assisted (Fig. \ref{fig:RIS-assisted System Topology Primary in 2-D}) and the second as RIS-primary (Fig. \ref{fig:RIS-primary System Topology Primary in 2-D}). In our previous work \cite{PACET_2024}, the RIS-assisted topology was studied in a beamforming scenario under both near and far field conditions and the main results were that the near field implementation should be preferred and that this topology does not provide sufficient improvement. Hence, following the direction of near field implementation, we are going to study RIS-primary topology and it will be shown that this topology can address the occurred issues of RIS-assisted topology and deal with double fading constraint. This improvement in performance is related with the collaboratively effective use of the RIS and VSAT directivity and the looser angular restrictions. More precisely, in the RIS-assisted case, there is a wider portfolio of possible orientations of RIS than the RIS-assisted case.

We are going to compare the received power of RIS-assisted and RIS-primary cases with the received power in case of a topology, where the VSAT is oriented towards the satellite and the RIS is absent. The improvement factor will be the percentage of improvement as follows:
\begin{equation}
    \text{Improvement Power Ratio} = \left(\frac{\frac{P_r}{P_t}|_{\text{with RIS}}^{\text{case}}}{\frac{P_r}{P_t}|_{\text{without RIS}}} - 1 \right) \cdot 100
\end{equation}
where $\text{case} = \{\text{RIS-assisted},\text{RIS-primary}\}$, 
\begin{equation}
    \frac{P_r}{P_t}|_{\text{with RIS}}^{\text{case}} = |h_{refl}^{beam} + h_{dir}^{beam}|^2
\end{equation} 
and 
\begin{equation}
    \frac{P_r}{P_t}|_{\text{without RIS}} = |h_{dir}^{beam}|^2|_{\text{without RIS}}
\end{equation}
The component $\frac{P_r}{P_t}|_{\text{without RIS}}$ refers to the case, in which the VSAT is oriented towards the satellite and the RIS is absent from the topology.

In RIS-primary topology, the VSAT is located in the center of the GCS, has direction of the beam the $Ox$ axis in its LCS and the satellite is located in the $xz$ plane. The LCS-RIS will be optimization parameters of this problem: $\text{center}_{\text{LCS}}^{\text{RIS}} = [x_{ris},y_{ris},z_{ris}]$ ,  $\text{rotation}_{\text{LCS}}^{\text{RIS}} = [\alpha_{ris},\beta_{ris},\gamma_{ris}]$. It should be mentioned that the direction of RIS beam is the $Ox$ axis in its LCS. The center of the LCS-VSAT does not change but because the VSAT should be focused on the center of LCS-RIS, the rotation vector of LCS-VSAT $\text{rotation}_{\text{LCS}}^{\text{vsat}} = [\alpha_{vsat},\beta_{vsat},\gamma_{vsat}]$ is also an optimization parameter. However, the RIS in this case is located in the directive axis of the VSAT, the position of it in spherical coordination in LCS-VSAT will be 
\begin{equation}
    \text{center}_{\text{LCS}}^{\text{RIS}} = [r = d_{ris}^{vsat},\theta = 90^{\circ}, \phi = 0^{\circ}]
\end{equation}
In addition, in order to mitigate the double fading constraint of RIS and because path loss attenuation has more impact to the power ratio than the antenna gain of the RIS element, the distance between RIS and the VSAT $d_{ris}^{vsat}$ is assumed to be equal to its down limit $R_\text{far field}^\text{parabolic}$. Furthermore, the topology has as axis of symmetry the imaginary direct line between the VSAT and satellite (cylindrical symmetry) and so we can study the system at a half plane with its boundary being this symmetry line. With this observation the optimization problem is reduced from the three dimensions to the two and we choose to solve the problem to the plain, which is defined through the rotation vector of LCS-VSAT as follows: \begin{equation}
    [\alpha_{vsat} = 180^{\circ},-180^{\circ}+\phi_{\alpha} \leq \beta_{vsat} \leq +\phi_{\alpha},\gamma_{vsat}]
\end{equation}
or equivalent 
\begin{equation}
    [\alpha_{vsat} = 0^{\circ},-180^{\circ}-\phi_{\alpha} \leq \beta_{vsat} \leq -\phi_{\alpha},\gamma_{vsat}]
\end{equation}
Further, the directional axis of RIS and the VSAT should be aligned in this topology and for this reason $\alpha_{ris} = 180 + \alpha_{vsat}$. Hence, the total optimization variables of the studied problem are $\left(  \beta_{ris},\gamma_{ris},\beta_{vsat},\gamma_{vsat} \right)$. In particular, $\gamma$ rotation parameter is responsible for the polarization of the VSAT and RIS due to their axis of propagation, $\beta_{ris}$ is responsible for maximizing the product $G_{RIS,(m,n)}^{sat} G_{RIS,(m,n)}^{vsat}$,because it controls the input and output angle of RIS, and $\beta_{vsat}$ determines the $d_{sat}^{(m,n)}$. As we can see in Fig. \ref{fig:RIS-primary System Topology Primary in 2-D}, the maximization of the product $G_{RIS,(m,n)}^{sat} G_{RIS,(m,n)}^{vsat}$ and the minimization of $d_{sat}^{(m,n)}$ are two contradictory features. 

\begin{equation}
    \begin{aligned}
        \min_{\beta_{ris}.\gamma_{ris},\beta_{vsat},\gamma_{vsat}} \quad & - |h_{refl}^{beam} + h_{dir}^{beam}|^2\\
        \text{s.t.} \quad & -180^{\circ}+\phi_{\alpha} \leq \beta_{vsat} \leq +\phi_{\alpha} \\
        \quad & -90^{\circ} \leq \beta_{ris} \leq + 90^{\circ} \\
        \quad & 0^{\circ} \leq \gamma_{vsat},\;\gamma_{ris} \leq + 360^{\circ}
    \end{aligned}
\end{equation}

The minimization optimization problem is not convex and as a result we are solving this problem numerically following a Successive Convex Approximation (SCA) method, \cite{Boyd_Convex_Optimization, Scutari_SCA ,SCA_Scutari_Theory_Part1}. In a nutshell, in SCA we approximate the concave (convex) or linear parts of a function, which we want to minimize (maximize), with a surrogate function and we solve the problem iteratively using the knowledge of the previous iteration results until convergence. The main construction rules for the surrogate function in terms of a minimization problem are the follows:
\begin{enumerate}
    \item $\Tilde{f}(\mathbf{x} | \mathbf{x}^{(k)})$ should be strongly convex in $\mathcal{X}$.
    \item $\nabla \Tilde{f}(\mathbf{x}^{(k)} | \mathbf{x}^{(k)}) = \nabla f(\mathbf{x}^{(k)}), \; \forall \; \mathbf{x}^{(k)} \in \mathcal{X}$
    \item $\nabla\Tilde{f}(\mathbf{x} | \mathbf{x}^{(k)})$ is continuous on $\mathcal{X}$
    \item \{Optional for further convergence results\} $\nabla \Tilde{f}(\mathbf{x} | \mathbf{x}^{(k)})$ is uniform Lipschitz continuous on $\mathcal{X}$
\end{enumerate}
where $\mathcal{X}$ is the domain of $\mathbf{x}$ and $\Tilde{f}(\mathbf{x} | \mathbf{x}^{(k)})$ is the surrogate function of the original function $f(\mathbf{x})$ in the $(k+1)$-th iteration. In our optimization problem more precisely, we approximate the non-convex function 
\begin{equation}
    F(\mathbf{x}) = - |h_{refl}^{beam}(\mathbf{x}) + h_{dir}^{beam}(\mathbf{x})|^2
\end{equation}
where $\mathbf{x} = [\beta_{vsat}, \beta_{ris}, \gamma_{vsat}, \gamma_{ris}]$ in the domain
\begin{equation} 
    \begin{aligned}
        \mathcal{X} = \{-180^{\circ}+\phi_{\alpha} \leq \beta_{vsat} \leq +\phi_{\alpha}\} \times \\ \{-90^{\circ} \leq \beta_{ris} \leq + 90^{\circ}\} \times \\
        \{0^{\circ} \leq \gamma_{ris} \leq + 360^{\circ}\} \times \\ \{0^{\circ} \leq \gamma_{vsat} \leq + 360^{\circ}\} 
    \end{aligned}
\end{equation}
with a Newton-based surrogate in each iteration $k = 0,1, \dots$ as follows: 
\begin{equation}
    \begin{aligned}
        \Tilde{F} \left( \mathbf{x} | \mathbf{x}^k \right) = F(\mathbf{x}^k) + \nabla F(\mathbf{x}^k)^T \cdot (\mathbf{x} - \mathbf{x}^k) +  \\\frac{\mu}{2} (\mathbf{x} - \mathbf{x}^k)^T \mathbf{H} (\mathbf{x} - \mathbf{x}^k)
    \end{aligned}
\end{equation}
The parameter $\mu$ is strictly positive and the matrix $\mathbf{H}$ is any positive definite matrix. For our case, we have chosen the matrix $\mathbf{H}$ to be the identity matrix. Hence, the surrogate function is strictly convex and the surrogate problem can be solved by using a gradient descent method or Lagrangian duality theory in each iteration $k = 0,1, \dots$. After each iteration $k$, we get an approximation of the optimization variables 
\begin{equation}
    \hat{\mathbf{x}} \left( \mathbf{x}^k \right) =  arg \min _ {\mathbf{x} \in \mathcal{X}} \Tilde{F} \left( \mathbf{x} | \mathbf{x}^k \right)
\end{equation}
and then we are going to update the value of $\mathbf{x}^{k+1}$ for the iteration $k+1$ as follows: 
\begin{equation}
    \mathbf{x}^{k+1} = \mathbf{x}^k + q^k \cdot \left( \hat{\mathbf{x}} \left( \mathbf{x}^k \right) - \mathbf{x}^k \right)
\end{equation} 
In order to have convergence of the iterative optimization algorithm, the parameter $q^k$ should follow the diminishing stepsize update formula, $q^{k+1} = q^k \cdot (1 - \epsilon q^k)$ ,$ q^0 < \frac{1}{\epsilon} \;,\; \epsilon \in (0,1)$ for every iteration $k = 0,1, \dots$. This choice along with the structure of the proposed surrogate function, which is uniformly Lipschitz continuous on $\mathcal{X}$, leads to the fact that all the limit points ${\mathbf{x}^k}$ are stationary, \cite{SCA_Scutari_Theory_Part1}.

\begin{algorithm}
\caption{SCA algorithm for optimization problem}\label{alg:sca}
\begin{algorithmic}
    \State Set $k=0$, initialize feasible point $\mathbf{x}^0 \in \mathcal{X}, \{q^k\} \in (0,1]$
    \Repeat
    \State $\hat{\mathbf{x}} \left( \mathbf{x}^k \right) =  arg \min _ {\mathbf{x} \in \mathcal{X}} \Tilde{F} \left( \mathbf{x} | \mathbf{x}^k \right)$
    \State $\mathbf{x}^{k+1} = \mathbf{x}^k + q^k \cdot \left( \hat{\mathbf{x}} \left( \mathbf{x}^k \right) - \mathbf{x}^k \right)$
    \State $k \leftarrow k + 1$
    \Until{convergence} \\
    \Return {$\mathbf{x}^k$}
\end{algorithmic}
\end{algorithm}

\section{Simulation Results}

\begin{figure}
    \centering
    \includegraphics[width=0.6\linewidth]{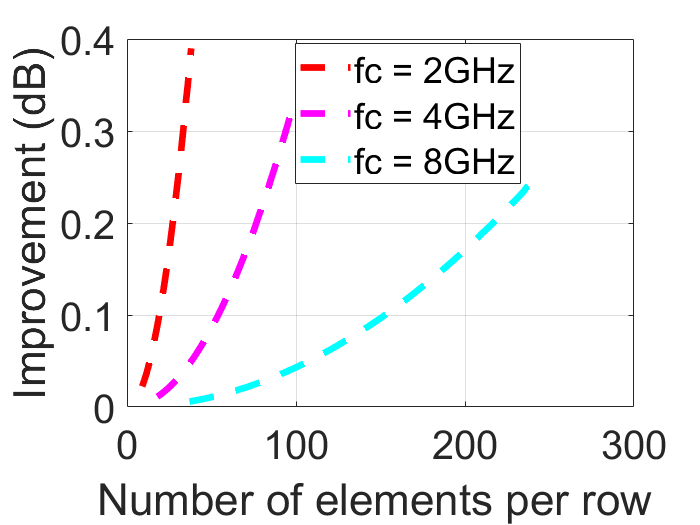}
    \caption{RIS-assisted, D = 1m.}
    \label{fig:RIS_assisted_1m}
\end{figure}
\begin{figure}
    \centering
    \includegraphics[width=0.6\linewidth]{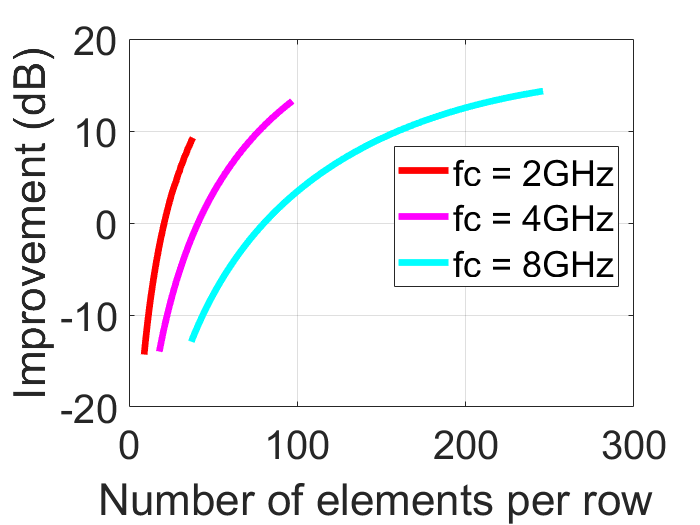}
    \caption{RIS-primary, D = 1m.}
    \label{fig:RIS_primary_1m}
\end{figure}
\begin{figure}
    \centering
    \includegraphics[width=0.6\linewidth]{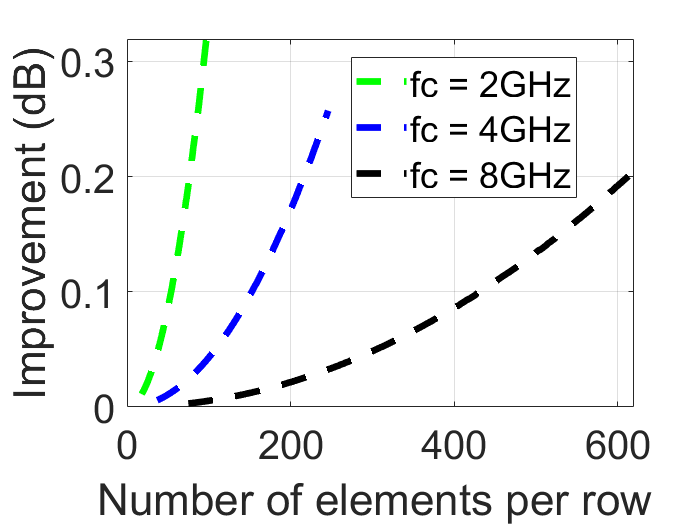}
    \caption{RIS-assisted, D = 2m.}
    \label{fig:RIS_assisted_2m}
\end{figure}
\begin{figure}
    \centering
    \includegraphics[width=0.6\linewidth]{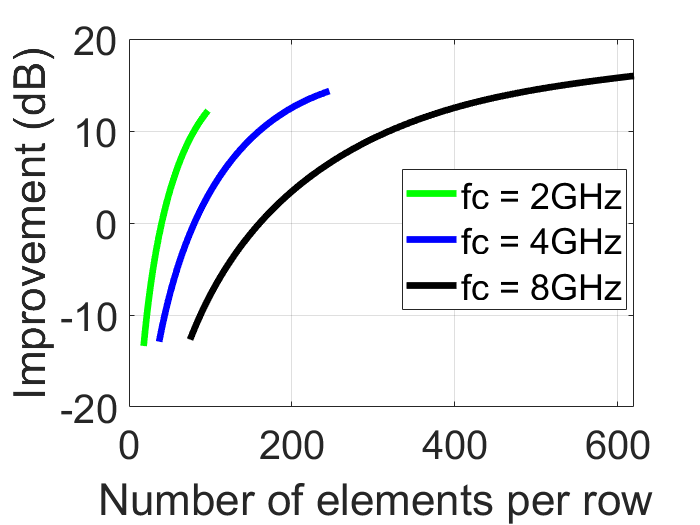}
    \caption{RIS-primary, D = 2m.}
    \label{fig:RIS_primary_2m}
\end{figure}

In this section we are simulating a hypothetical GEO satellite system at height of $38.000$ km from the ground which is located in $\phi_s = 39^{\circ} \text{E}$, which is the position of satellite HELLAS-SAT 2. The earth is considered as an ideal sphere with radius $R_e = $6371km and the VSAT is located at $(\phi_s = 37^{\circ}\text{E},\theta_s = 23^{\circ}\text{N})$, which is the location of NTUA campus at Zografou, Athens. Two different diameters for VSAT are considered, $D = \{1\text{m},2\text{m}\}$ and two different downlink frequency operating bands: S at 2 GHz, C at 4 GHz. Finally, we should mention that we consider horizontal polarization of electromagnetic signals, which depends significantly on the orientation of the antennas. The spacing of the RIS will be equal to $d_x = d_y = \frac{\lambda}{2}$, the number of rows $M$ is equal to the number of columns $N$, all elements are passive ($A_{m,n} = 1$) and the distance between RIS and the VSAT is equal to $R_\text{far field}^\text{parabolic}$. As we referred to previously, the VSAT is located at the radiating near field of RIS and we are going to evaluate the improvement provided by using RIS as a function of VSAT diameter, carrier frequency and number of RIS elements per row, under the constraint of equation (\ref{eq: Freznel Zone Bound}), which describes the range of number of elements per dimension of a square RIS in order that the receiver (VSAT in this case) is in the radiating near field of it.

As easily observed from the diagrams in Fig. \ref{fig:RIS_assisted_1m} - \ref{fig:RIS_primary_2m}, the RIS-primary topology demonstrates greater enhancements compared to the RIS-assisted topology. This occurs because we capitalize on the directivity of VSAT antennas to equalize the dual fading constraints of the reflective path, taking into consideration the VSAT position in the radiating near field of the RIS. In the case of RIS-assisted topology, the RIS is served from the lower side lobes of the VSAT, resulting in marginal improvement. It is crucial to emphasize that the negative improvement in decibel scale in the RIS-primary topology can be interpreted as follows: the RIS gain, approximately equal to $A_{m,n} \cdot M \cdot N$, is not sufficiently high to mitigate the dual fading. Furthermore, we observe that with an increase in frequency and under the same VSAT diameter, the improvement decreases in RIS-assisted topology and increases in RIS-primary topology.

Further, it is interesting to compare different use cases under the same topology. Firstly, for the same VSAT diameter, as the frequency increases, more elements per dimension of the RIS are required to achieve the same improvement. For example, to achieve a 3dB improvement (doubling of the power margin) in the RIS-primary case with a VSAT diameter of 1m, we require 25 elements per dimension at 2 GHz, 49 elements per dimension at 4 GHz, and 97 elements per dimension at 8 GHz. However, these requirements translate to the need for a RIS with a length approximately equal to 1.85m for all three cases. Furthermore, for the same carrier frequency, as the diameter of the VSAT increases, higher maximum improvements can be achieved, but simultaneously, more elements per dimension of the RIS are needed to achieve the same improvement as expected. Hence, it is evident that there is a practical limitation in this case as we need an equal size RIS in order to improve the performance of the classic VSAT system.

\section*{Conclusion}
In conclusion, it is evident that the implementation of RIS can provide sufficient improvement to the power margin of a FSS system under the condition of optimum positioning of RIS. Moreover, it is necessary that the VSAT is located at the near field of RIS and to focus its antenna on RIS. An important outcome of the simulation was the decrease of the improvement as the frequency increases and the number of RIS elements does not change. In addition, it should be mentioned that the size of RIS must be of the order of magnitude of diameter VSAT in order to achieve a notable improvement to power margin, for instance 3 dB. A future research step will be the study of interference, which is caused by this system topology, when many pairs of VSAT-RIS, which maybe are linked with different satellites, are located near to each other.

\bibliographystyle{IEEEtran}
\bibliography{my_bibliography_RIS_SatCom}

@ARTICLE{RIS_Survey_1,
  title     = "Reconfigurable {I}ntelligent {S}urfaces for {W}ireless {C}ommunications: Principles, {C}hallenges, and {O}pportunities",
  author    = "ElMossallamy, Mohamed A. and Zhang, Hongliang and Song, Lingyang and Seddik, Karim G. and Han, Zhu and Li, Geoffrey Ye",
  journal   = "IEEE Transactions on Cognitive Communications and Networking",
  publisher = "Institute of Electrical and Electronics Engineers (IEEE)",
  volume    =  "6",
  number    =  "3",
  pages     = "990-1002",
  month     =  "Sept.",
  year      =  "2020",
  doi       = "10.1109/TCCN.2020.2992604"
}

@ARTICLE{RIS_Survey_2,
  title     = "Reconfigurable {I}ntelligent {S}urfaces for {6G} {S}ystems: Principles, {A}pplications, and {R}esearch {D}irections",
  author    = "Pan, Cunhua and Ren, Hong and Wang, Kezhi and Kolb, Jonas Florentin and Elkashlan, Maged and Chen, Ming and Di Renzo, Marco and Hao, Yang and Wang, Jiangzhou and Swindlehurst, A. Lee and You, Xiaohu and Hanzo, Lajos",
  journal   = "IEEE Communications Magazine",
  publisher = "Institute of Electrical and Electronics Engineers (IEEE)",
  volume    =  "59",
  number    =  "6",
  pages     = "14-20",
  month     =  "June",
  year      =  "2021",
  doi       = "10.1109/MCOM.001.2001076"
}

@ARTICLE{RIS_Satellite_Survey_1,
  title     = "Evolution of {N}on-{T}errestrial {N}etworks {F}rom {5G} to {6G}: A {S}urvey",
  author    = "Azari, M. Mahdi and Solanki, Sourabh and Chatzinotas, Symeon and Kodheli, Oltjon and Sallouha, Hazem and Colpaert, Achiel and Mendoza Montoya, Jesus Fabian and Pollin, Sofie and Haqiqatnejad, Alireza and Mostaani, Arsham and Lagunas, Eva and Ottersten, Bjorn",
  journal   = "IEEE Communications Surveys and Tutorials",
  publisher = "Institute of Electrical and Electronics Engineers (IEEE)",
  volume    =  "24",
  number    =  "4",
  pages     = "2633-2672",
  year      =  "2022",
  doi       = "10.1109/COMST.2022.3199901"
}

@ARTICLE{RIS_Satellite_Survey_2,
  title     = "Nonterrestrial {C}ommunications {A}ssisted by {R}econfigurable {I}ntelligent {S}urfaces",
  author    = "Ye, Jia and Qiao, Jingping and Kammoun, Abla and Alouini,
               Mohamed-Slim",
  journal   = "Proceedings of the IEEE",
  publisher = "Institute of Electrical and Electronics Engineers (IEEE)",
  volume    =  "110",
  number    =  "9",
  pages     = "1423-1465",
  month     =  "Sept.",
  year      =  "2022",
  doi       = "10.1109/JPROC.2022.3169690"
}

@ARTICLE{RIS_Satellite_Hopping,
  title     = "Reconfigurable {I}ntelligent {S}urfaces {E}mpowered {THz}
               {C}ommunication in {LEO} {S}atellite {N}etworks",
  author    = "Tekbiyik, Kursat and Kurt, Gunes Karabulut and Ekti, Ali Riza and Yanikomeroglu, Halim",
  journal   = "IEEE Access",
  publisher = "Institute of Electrical and Electronics Engineers (IEEE)",
  volume    =  "10",
  pages     = "121957-121969",
  year      =  "2022",
  doi       = "10.1109/ACCESS.2022.3223086"
}

@ARTICLE{RIS_Satellite_NLoS,
  title        = "Enabling {NLoS} {LEO} {S}atellite {C}ommunications with
                  {R}econfigurable {I}ntelligent {S}urfaces",
  author       = "Xiaowen Tian and Nuria Gonzalez-Prelcic and Takayuki Shimizu",
  year         =  "2022",
  eprint="2205.15528",
  archivePrefix="arXiv",
  primaryClass="eess.SP",
  url="https://arxiv.org/abs/2205.15528"
}

@ARTICLE{RIS_Satellite_Refracting,
  title     = "Refracting {RIS-aided} {H}ybrid {S}atellite-{T}errestrial {R}elay {N}etworks: Joint {B}eamforming {D}esign and {O}ptimization",
  author    = "Lin, Zhi and Niu, Hehao and An, Kang and Wang, Yong and Zheng, Gan and Chatzinotas, Symeon and Hu, Yihua",
  journal   = "IEEE Transactions on Aerospace and Electronic Systems",
  publisher = "Institute of Electrical and Electronics Engineers (IEEE)",
  volume    =  "58",
  number    =  "4",
  pages     = "3717--3724",
  month     =  "Aug.",
  year      =  "2022",
  doi       = "10.1109/TAES.2022.3155711"
}

@ARTICLE{RIS_Orientation_Optimization,
  title     = "Orientation {O}ptimization {W}ith {L}ow-{B}it {D}iscrete {P}hase {S}hifts for {RIS}-{A}ided {C}ommunication",
  author    = "Yang, Lei and Shen, Shengqiang and Ma, Shuai and Xu, Gang and Li, Shiyin",
  journal   = "IEEE Wireless Communications Letters",
  publisher = "Institute of Electrical and Electronics Engineers (IEEE)",
  volume    =  "12",
  number    =  "12",
  pages     = "1983-1987",
  month     =  "Dec.",
  year      =  "2023",
  doi       = "10.1109/LWC.2023.3289985"
}

@ARTICLE{RIS_Sat_Near_Field,
  title     = "{E}nergy-{E}fficient {RIS}-{A}ssisted {S}atellites for {I}o{T} {N}etworks",
  author    = "Tekbiyik, Kursat and Kurt, Gunes Karabulut and Yanikomeroglu,Halim",
  journal   = "IEEE Internet of Things Journal",
  publisher = "Institute of Electrical and Electronics Engineers (IEEE)",
  volume    =  "9",
  number    =  "16",
  pages     = "14891--14899",
  month     =  "Aug.",
  year      =  "2022",
  doi       = "10.1109/JIOT.2021.3112881"
}

@ARTICLE{RIS_LEO_Mobility,
  title     = "Intelligent {R}eflecting {S}urface {O}peration {U}nder {P}redictable {R}eceiver {M}obility: A {C}ontinuous {T}ime {P}ropagation {M}odel",
  author    = "Matthiesen, Bho and Bjornson, Emil and De Carvalho, Elisabeth and Popovski, Petar",
  journal   = "IEEE Wireless Communications Letters",
  publisher = "Institute of Electrical and Electronics Engineers (IEEE)",
  volume    =  "10",
  number    =  "2",
  pages     = "216--220",
  month     =  "Feb.",
  year      =  "2021",
  doi       = "10.1109/LWC.2020.3024781"
}

@INPROCEEDINGS{RIS_Orinetation_Satellite,
  author={Zheng, Ziyuan and Jing, Wenpeng and Lu, Zhaoming and Wen, Xiangming},
  booktitle={2022 IEEE Globecom Workshops (GC Wkshps)}, 
  title={RIS-{E}nhanced {LEO} {S}atellite {C}ommunication: {J}oint {P}assive {B}eamforming and {O}rientation {O}ptimization}, 
  year={2022},
  volume={},
  number={},
  pages={874-879},
  doi={10.1109/GCWkshps56602.2022.10008708}}

@techreport{ITU_R_FSS,
     title = "Reference radiation pattern for earth station antennas in the fixed-satellite service for use in coordination and interference assessment in the frequency range
     from 2 to 31 GHz",
     year = {2010},
     institution = {ITU-R S.465-6},
     month = {01},

}

@techreport{FSS_Radiation_Pattern,
     title = "Methodology on the modelling of earth station antenna gain in the region of
     the antenna main-lobe and the transition region between the minimum
     angle of the reference antenna pattern and the main-lobe",
     year = {2010},
     institution = {ITU-R S.2196},
     month = {07},

}

@BOOK{Ha,
  title     = "Digital Satellite Communications",
  author    = "Ha, Tri T",
  publisher = "McGraw-Hill",
  edition   =  "2",
  month     =  "April",
  year      =  "1990",
  address   = "New York, NY"
}

@ARTICLE{Sak,
  title     = "Cooperative {H}ybrid {L}and {M}obile {S}atellite-{T}errestrial {B}roadcasting {S}ystems: {O}utage {P}robability {E}valuation and {A}ccurate {S}imulation",
  author    = "Sakarellos, Vassileios K and Kourogiorgas, Charilaos and
               Panagopoulos, Athanasios D",
  journal   = "Wireless Personal Communications",
  publisher = "Springer Science and Business Media LLC",
  volume    =  "79",
  number    =  "2",
  pages     = "1471-1481",
  month     =  "Nov.",
  year      =  "2014",
  doi       = "10.1007/s11277-014-1941-6"
}

@techreport{3GPP_5G_channel_modeling,
     title = "{5G}, study on channel model for frequencies from 0.5 to 100 {GH}z (3{GPP} {TR} 38.901 Version 16.1.0 {R}elease 16)",
     year = {2020},
     institution = {ETSI TR 138 901 v16.1.0},
     month = {11},

}

@ARTICLE{RIS_Channel_Modeling_1,
  title     = "Wireless {C}ommunications {W}ith {R}econfigurable {I}ntelligent {S}urface: Path {L}oss {M}odeling and {E}xperimental {M}easurement",
  author    = "Tang, Wankai and Chen, Ming Zheng and Chen, Xiangyu and Dai, Jun Yan and Han, Yu and Di Renzo, Marco and Zeng, Yong and Jin, Shi and Cheng, Qiang and Cui, Tie Jun",
  journal   = "IEEE Transactions on Wireless Communications",
  publisher = "Institute of Electrical and Electronics Engineers (IEEE)",
  volume    =  "20",
  number    =  "1",
  pages     = "421--439",
  month     =  "Jan.",
  year      =  "2021",
  doi       = "10.1109/TWC.2020.3024887"
}

@techreport{rain_attenuation,
     title = "Specific attenuation model for rain for use in prediction methods",
     year = {2005},
     institution = {ITU-R P.838-3},
     month = {03},

}

@techreport{gases_attenuation,
     title = "Attenuation by atmospheric gases",
     year = {2022},
     institution = {ITU-R P.676-13},
     month = {08},

}

@techreport{earth_space_propagation,
     title = "Propagation data and prediction methods required for the design of Earth-space telecommunication systems",
     year = {2023},
     institution = {ITU-R P.618-14},
     month = {08},

}

@ARTICLE{PACET_2024,
  title     = "Enhancing {S}atellite {C}ommunications {C}onnectivity with
               {R}econfigurable {I}ntelligent {S}urfaces : {R}eflection {L}ink {O}ptimization",
  journal = "2024 Panhellenic Conference on Electronics \& Telecommunications ({PACET})",
  author    = "Vaidanis, Symeon and Panagopoulos, Athanasios D",
  year={2024},
  pages     = "1-5",
  doi       = "10.1109/PACET60398.2024.10497063"
}

@book{Boyd_Convex_Optimization,
    author = "Boyd, Stephen and Vandenberghe, Lieven ",
    title = "Convex {O}ptimization",
    publisher = "Cambridge University Press",
    year = "2004",
    doi = "10.1017/CBO9780511804441"
}

@Inbook{Scutari_SCA,
    author="Scutari, Gesualdo and Sun, Ying",
    title="Parallel and {D}istributed {S}uccessive {C}onvex {A}pproximation {M}ethods for {B}ig-{D}ata {O}ptimization",
    bookTitle="Multi-agent Optimization: Cetraro, Italy 2014",
    year="2018",
    publisher="Springer International Publishing",
    address="Cham",
    pages="141-308",
    doi="10.1007/978-3-319-97142-1_3",
    url="https://doi.org/10.1007/978-3-319-97142-1_3"
}

@misc{SCA_Scutari_Theory_Part1,
  author={Scutari, Gesualdo and Facchinei, Francisco and Lampariello, Lorenzo},
  journal={IEEE Trans. on Signal Processing}, 
  title={Parallel and Distributed Methods for Constrained Nonconvex Optimization—Part I: Theory}, 
  year={2017},
  volume={65},
  number={8},
  pages={1929-1944},
  doi={10.1109/TSP.2016.2637317}}

@ARTICLE{RIS_LEO_GEO,
  author={Zheng, Ziyuan and Jing, Wenpeng and Lu, Zhaoming and Wu, Qingqing and Zhang, Haijun and Gesbert, David},
  journal={IEEE Journal on Selected Areas in Communications}, 
  title={Cooperative {M}ulti-{S}atellite and {M}ulti-{RIS} {B}eamforming: {E}nhancing {LEO} {S}at{C}om and {M}itigating {LEO-GEO} {I}ntersystem {I}nterference}, 
  year={2024},
  volume={},
  number={},
  pages={1-1},
  doi={10.1109/JSAC.2024.3460068}
}

@ARTICLE{RIS_Multisatellite,
  author={Feng, Kai and Zhou, Ting and Xu, Tianheng and Chen, Xianfu and Hu, Honglin and Wu, Celimuge},
  journal={IEEE Internet of Things Journal}, 
  title={Reconfigurable {I}ntelligent {S}urface-{A}ssisted {M}ultisatellite {C}ooperative {D}ownlink {B}eamforming}, 
  year={2024},
  volume={11},
  number={13},
  pages={23222-23235},
  doi={10.1109/JIOT.2024.3372091}
}

@ARTICLE{RIS_Satellite_NOMA,
  author={Lv, Ziwei and Gong, Fengkui and Chen, Gaojie and Li, Guo and Hui, Tengfei and Xu, Sai},
  journal={IEEE Wireless Communications Letters}, 
  title={Energy {E}fficiency {D}esign in {RIS}-{A}ssisted {S}atellite–{T}errestrial {I}ntegrated {N}etworks {W}ith {NOMA}}, 
  year={2024},
  volume={13},
  number={7},
  pages={1948-1952},
  doi={10.1109/LWC.2024.3397347}
}

@ARTICLE{RIS_Multicast_Satellite,
  author={Kim, Taehyoung and Jung, Minchae and Son, Hyukmin},
  journal={IEEE Transactions on Aerospace and Electronic Systems}, 
  title={Joint {U}ser {S}cheduling and {P}hase {S}hift {O}ptimization for {STAR-RIS-A}ssisted {M}ulticast {S}atellite {C}ommunications}, 
  year={2024},
  volume={60},
  number={5},
  pages={7466-7474},
  doi={10.1109/TAES.2024.3371398}
}

\end{document}